\documentclass[12pt]{iopart}

\renewcommand{\vec}[1]{{\bf {#1}}}
\begin{document}

\title[]{The Partition Function in the Wigner-Kirkwood expansion}

\author{Sergei G.~Matinyan$^1$ 
\footnote[1]{Present address: 3106 Hornbuckle Place, Durham, NC 27707, USA.}
and Berndt M\"uller$^2$}

\address{$^1$ Yerevan Physics Institute, 375036 Yerevan, Armenia}

\address{$^2$ Department of Physics, Duke University, Durham, NC 27708}

\date{\today}

\begin{abstract}
We study the semiclassical Wigner-Kirkwood (WK) expansion of the
partition function $Z(t)$ for arbitrary even homogeneous potentials, 
starting from the Bloch equation. As is well known, the phase-space kernel
of $Z$ satisfies the so-called Uhlenbeck-Beth equation, which depends on 
the gradients of the potential. We perform a chain of transformations to
obtain novel forms of this equation that invite analogies with various 
physical phenomena and formalisms, such as diffusion processes, the 
Fokker-Planck equation, and supersymmetric quantum mechanics.
\end{abstract}

\section{Introduction}

The problem of the semiclassical expansion of the physical states and 
their energies in power of $\hbar$ rather than the expansions involving
oscillating functions of $\hbar^{-1}$ has a long history and was studied 
and used in different fields of science (see \cite{bib01} for review)
beginning from the Thomas-Fermi (TF) approximation \cite{bib01,bib02,bib03}.
Being the lowest order in $\hbar$, the TF correction to the classical 
partition function (heat kernel) takes into account the discreteness 
of the phase-space treating the Hamiltonian classically.

We consider here Hamiltonians of the form $H = T + V$ where $T$ is
the kinetic and $V$ the potential energy, respectively. The 
non-commutativity of the $T$ and $V$ in the quantum Hamiltonian
\begin{equation}
\hat H = -\frac{\hbar^2}{2}\nabla^2 + V(\{x_i\}) 
\label{eq01}
\end{equation}
where $\nabla^2$ denotes the Laplacian and $i$ labels the coordinates,
was taken into account by the Wigner-Kirkwood (WK) approach 
\cite{bib04,bib05}. (Note: We set the mass $m=1$, and all quantities 
are given in units of energy $E$ (implying the dimensions: $[H]=1, 
[t]=-1, [\hbar] = 3/4, [V]=1, [x]=1/4)$.) We do not specify the form 
of the, generally non-central, potential $V(\vec{x})$ for $n$ degrees
of freedom ($\vec{x} = (x_1,x_2,\ldots,x_n)$). 
The case of even homogeneous polynomials satisfying 
$V(\lambda \vec{x}) = \lambda^{2N} V(\vec{x})$ includes the potentials
of the well-known Yang-Mills classical and quantum mechanics 
\cite{bib07,bib08,bib09} (see \cite{bib10} for a review). 

\section{The Partition Function in the WK expansion}

The partition function
\begin{equation}
Z(t) = {\rm Tr} \left[\exp(-t{\hat H})\right] 
= \sum_{n=0}^{\infty} e^{-E_nt},
\label{eq02}
\end{equation}
which is the Laplace transform of the density of states ${\cal L}[\rho(E)]$, 
is calculated taking the Wigner transform of the quantum operator
$\exp(-t\hat H)$ and integrating it over the phase-space with measure
\begin{equation}
d\Gamma = (2\pi\hbar)^{-n} d^nx d^np,
\nonumber 
\end{equation}
where $n$ is the dimension of the system.

For the semiclassical expansion it is convenient to use the plane waves
as a complete set:
\begin{equation}
Z(t) = (2\pi\hbar)^{-n} \int d\Gamma
e^{-i\vec{p}\cdot\vec{x}/\hbar}\, e^{-t\hat H} \,
e^{i\vec{p}\cdot\vec{x}/\hbar} .
\label{eq03}
\end{equation}
The WK method takes into account only part of the quantum features, 
but ignors the quantum fluctuations defined by the concrete form of 
the potential and dynamics (see \cite{bib09} for the discussion of 
the role of the quantum fluctuations in Yang-Mills quantum mechanics).
Following \cite{bib06}, we can account the non-commuting terms in WK 
approach setting
\begin{equation}
e^{-t\hat H} \, e^{i\vec{p}\cdot\vec{x}/\hbar}
= e^{-tH} e^{i\vec{p}\cdot\vec{x}/\hbar}\, W(\vec{x},\vec{p};t)
= u(\vec{x},\vec{p};t) ,
\label{eq04}
\end{equation}
where $H$ is the classical Hamiltonian and the function 
$W (\vec{x},\vec{p};t)$ is to be determined.
The function $u(\vec{x},\vec{p};t)$ satisfies the Bloch equation, 
the analogue of the time dependent Schr\"odinger equation:
\begin{equation}
\frac{\partial u}{\partial t} + {\hat H} u = 0
\label{eq05}
\end{equation}
with the initial condition
\begin{equation}
\lim_{t\to 0} u(\vec{x},\vec{p};0) = e^{i\vec{p}\cdot\vec{x}/\hbar} ,
\label{eq06}
\end{equation}
corresponding to the initial condition $W(\vec{x},\vec{p};0) = 1$.

Boundary conditions on $u(\vec{x},\vec{p};t)$ at $|x| = \infty$ are in 
correspondence with the boundary conditions on $W(\vec{x},\vec{p};t)$.
From (\ref{eq04}) and (\ref{eq05}) we obtain an exact equation for 
$W(\vec{x},\vec{p};t)$, which we call the Uhlenbeck-Beth (UB) equation):
\begin{eqnarray}
\frac{\partial W}{\partial t} = \frac{\hbar^2}{2}
\left[ \nabla^2 - t(\nabla^2V) + t^2(\nabla V)^2 - 2t(\nabla V)\cdot\nabla
\right. \nonumber \\
\left. \qquad\qquad
+\frac{2i}{\hbar}\vec{p}\cdot(\nabla - t\nabla V) \right] W .
\label{eq07}
\end{eqnarray}
Next we expand $W$ in powers of $\hbar$ \cite{bib05}:
\begin{equation}
W = \sum_{k=0}^\infty \hbar^k W_k 
\label{eq08}
\end{equation}
and equate the terms with the same power of $\hbar$ on the both sides.
The result is:
\begin{eqnarray}
\frac{\partial W_k}{\partial t} = \frac{1}{2}
\left[ \nabla^2 - t(\nabla^2V) + t^2(\nabla V)^2 - 2t(\nabla V)\cdot\nabla
\right] W_{k-2} 
\nonumber \\ \qquad\qquad
+ i\vec{p}\cdot(\nabla - t\nabla V) W_{k-1} .
\label{eq09}
\end{eqnarray}
with conditions $W_0 = 1, W_k = 0$ for $k<0$.

\section{New forms of the UB equation}

The expressions (\ref{eq07}) and (\ref{eq09}) can be written in more 
compact form, if we introduce the ``covariant'' derivative
$\vec{\cal D} = \nabla - t(\nabla V)$:
\begin{equation}
\frac{\partial W}{\partial t} 
= \frac{\hbar^2}{2} \left[ {\vec{\cal D}}^2 
  + \frac{2i}{\hbar}\vec{p}\cdot\vec{\cal D} \right] W 
= \frac{1}{2} \left[ (\hbar\vec{\cal D} + i\vec{p})^2 
  + \vec{p}^2 \right] W .
\label{eq10}
\end{equation}
or its recursive form
\begin{equation}
\frac{\partial W_k}{\partial t} 
= \frac{1}{2}\left[ {\vec{\cal D}}^2 W_{k-2} + 2i\vec{p}\cdot{\cal D} 
  W_{k-1} \right] .
\label{eq11}
\end{equation}
We note that the symbol $\vec{p}$ in (\ref{eq10},\ref{eq11}) denotes a 
classical phase-space variable and not an operator. To our knowledge,
the forms (\ref{eq10}) and (\ref{eq11}) of eqs.~(\ref{eq07}) and 
(\ref{eq09}) are novel and have not been given in the literature.

In terms of the operator $\vec{\cal D}$, eq.~(\ref{eq10}) resembles a 
Fokker-Planck equation (FPE) for $W(\vec{x},\vec{p};t)$ with the 
diffusion constant $\hbar^2$ and the constant drift vector $-i\hbar\vec{p}$.
The relation to the FPE can be further elucidated by noting that the 
``vector potential'' $\vec{A} = t\nabla V$ is a complete gradient and 
thus may be ``gauged out'' by the transformation $W \to \exp(tV)W'$, 
yielding an alternative form of (\ref{eq10}):
\begin{equation}
\frac{\partial W'}{\partial t} 
= \frac{\hbar^2}{2} \left[ \nabla^2 + \frac{2i}{\hbar}\vec{p}\cdot\nabla 
  \right] W' - VW' .
\label{eq12}
\end{equation}
If we interpret $W'(\vec{x},\vec{p};t)$ as an one-time probability density 
and introduce the probability current (sometimes called the probability
flux in the literature \cite{bib11,bib12}):
\begin{equation}
\vec{J} = -i\hbar\vec{p}W' - \frac{\hbar^2}{2}\nabla W'
= W'\left( -i\hbar\vec{p} - \frac{\hbar^2}{2}\nabla \ln W' \right) ,
\label{eq13}
\end{equation}
we may write (\ref{eq12}) in the form of the continuity equation
\begin{equation}
\frac{\partial W'}{\partial t} + \nabla\cdot\vec{J}'
= -V(x) W'(\vec{x},\vec{p};t) ,
\label{eq14}
\end{equation}
where the potential term acts as a source term leading to a ``probability 
loss'' and violates the local conservation law associated with FPE. 
Integrating over volume and using Gauss' theorem in (\ref{eq14}) we obtain
\begin{equation}
\frac{\partial}{\partial t} \int W'\, d^nx
+ \int_{S} d{\vec\sigma}\cdot\vec{J}' = -\int V W'\, d^nx ,
\label{eq15}
\end{equation}
where $S$ is the surface confining the volume. If the flux vanishes at 
infinity we have
\begin{equation}
\frac{\partial}{\partial t} \int W'\, d^nx
= -\int V W'\, d^nx ,
\label{eq16}
\end{equation}
Note that in the right-hand side of (\ref{eq12}) and (\ref{eq14}) 
we may substitute the total energy $H$ instead of the potential $V$ by 
making the transformation $W'=W''\exp(-p^2t/2)$, since p commutes with 
gradients on the right-hand side of (12) and (14).

The presence of the potential term $VW'$ (or $UW'$) in (\ref{eq12}) and 
(\ref{eq14}) invalidates the standard result for the FPE (see, e.~g. 
\cite{bib11,bib12}) that under rather loose conditions on the drift and 
diffusion kernels (which are satisfied in our case) all solutions of 
the FPE must coincide for sufficiently large times. If drift and 
diffusion functions do not depend on time, stationary solution may 
exist, and one solution is unique in the sense that all 
other solutions agree with it after sufficiently long time.
Having the stationary solution of the FPE, one is able to find the 
general solution with given initial and boundary conditions (see, e.~g.
\cite{bib13,bib14}).

In our case it is clear that stationary solutions to the Bloch equation 
(\ref{eq05}) exist only for the trivial case of vanishing $V$. This is 
evident from (\ref{eq01}), (\ref{eq04}) and (\ref{eq05}): Using the 
second expression for the probability flux in (\ref{eq13}), setting 
${\dot W}_{\rm st}=0$ and using $U$ instead of $V$ in (\ref{eq14}), we obtain 
the equation
\begin{equation}
(\nabla \ln u_{\rm st})^2 = \frac{2}{\hbar^2} V(\vec{x}) .
\label{eq17}
\end{equation}
Since $u_{\rm st} = \exp(i\vec{p}\cdot\vec{x}/\hbar)$ is the 
solution of the Bloch equation (\ref{eq05}) 
with $\partial u/\partial t =0$, we arrive at the stated conclusion
that there are no stationary solutions to the Bloch equation for 
non-vanishing potential. Similarly, for the FPE (\ref{eq12}) resulting 
from the substitution $W = \exp(tV) W'$, it is clear from (\ref{eq04}) 
that the only stationary solution is $W'=1$ for $V(\vec{x}) = 0$.

Kepping this in mind, let us divide the solution 
to (\ref{eq12}) into a stationary and a time-dependent part:
\begin{equation}
W'(\vec{x},\vec{p};t) = W'_{\rm st}(\vec{x},\vec{p}) + w'(\vec{x},\vec{p};t).
\label{eq17a}
\end{equation}
For $V=0$ there exists the stationary solution
\begin{equation}
L_{\rm FP}(\vec{x},\vec{p}) W'_{\rm st} \equiv \frac{\hbar^2}{2} 
\left[\nabla^2 + \frac{2i}{\hbar}\vec{p}\cdot\nabla \right]
W'_{\rm st} = 0 .
\label{eq17b}
\end{equation}
Inserting (\ref{eq17a}) into (\ref{eq12}) we obtain:
\begin{equation}
\frac{\partial w'}{\partial t} = (L_{\rm FP}-V)w' - VW'_{\rm st} .
\label{eq17c}
\end{equation}
Following the standard approach to the solution of inhomogeneous
first-order differential equations, we can obtain a formal solution 
of (\ref{eq17c}) in the form
\begin{equation}
w'(\vec{x},\vec{p};t) = - \int_{-\infty}^t dt'\, 
  e^{[L_{\rm FP}-V](t-t')} V(\vec{x}) W'_{\rm st}(\vec{x},\vec{p}) .
\label{eq17d}
\end{equation}
To this we need to add the general solution of the homogeneous equation
\begin{equation}
\frac{\partial \tilde{w}}{\partial t} = (L_{\rm FP}-V)\tilde{w} .
\label{eq17e}
\end{equation}
The complete solution (\ref{eq17a}) thus has three contributions:
\begin{equation}
W'(\vec{x},\vec{p};t) = W'_{\rm st}(\vec{x},\vec{p}) + w'(\vec{x},\vec{p};t)
  + \sum_n C_n\tilde{w}_n(\vec{x},\vec{p};t) ,
\label{eq17f}
\end{equation}
where the $C_n$ are derived from the initial condition 
$W'(\vec{x},\vec{p};0)=1$. The solutions $\tilde{w}_n$ of the homogeneous 
equation are obtained by separation of variables, 
$\tilde{w}_n(\vec{x},\vec{p};t)=T_n(t)X_n(\vec{x},\vec{p})$, in the form
\begin{eqnarray}
T_n(t) = e^{-\lambda_n t},
\label{eq17g}
\\
(L-V)X_n(\vec{x},\vec{p}) = -\lambda_n^2 X_n(\vec{x},\vec{p}) ,
\label{eq17h}
\end{eqnarray}
where $-\lambda_n^2$ are the eigenvalues of the operator $L-V$.
Finally, the solution of (\ref{eq17b}) is the plane wave:
\begin{equation}
W'_{\rm st}(\vec{x},\vec{p}) 
= \frac{c}{\hbar^2} e^{-2i\vec{p}\cdot\vec{x}/\hbar} .
\label{eq17i}
\end{equation}

\section{Chain of transformations of the UB equation}

By the transformation $W = \exp(tV)W'$ (or $W = \exp(tU)W'$) we 
obtained the equation (\ref{eq12}) which is intermediate between 
the Bloch and UB equations. Going one step further by setting 
$W' = \exp(-i\vec{p}\cdot\vec{x}/\hbar){\tilde W}$ we arrive,
of course, at the Bloch equation (\ref{eq05}) as it is evident 
from the chain of transformations
\begin{equation}
W \to e^{tU}\,W'' \to e^{tU-i\vec{p}\cdot\vec{x}/\hbar}\,\tilde W ,
\label{eq18}
\end{equation}
with the identification $\tilde W = u$.) Each link of this chain 
opens up new analogies with the important physical processes and 
equations which govern them. We elaborate this issue below in more 
detail. Before we address this subject, however, we note that the 
expansion in $\hbar$ proposed by Kirkwood \cite{bib05} is also an
expansion in powers of the gradient operator as emphasized in 
ref.~\cite{bib06}: The gradient operator occurs $k$ times for each 
$W_k$ in eq.~(\ref{eq09}). In the general case, one needs to expand 
in powers of $\hbar$ or, equivalently, in powers of the gradient 
operator using (\ref{eq09}) and (\ref{eq11}), as done in 
\cite{bib08,bib09}.

In some special cases the compact form of (\ref{eq10}) or (\ref{eq12}) 
and the Bloch equation itself may be the starting point of another 
approximation scheme, which can bring to bear the knowledge of
approximation techniques developed for the Fokker-Planck and 
Schr\"odinger equations. With this in mind, we explore here several 
modified forms of the diffusion equation (\ref{eq12}), before we consider 
some aspects of the Bloch equation. Going back to our original function 
$W(\vec{x},\vec{p};t) = \exp(tU) W'(\vec{x},\vec{p};t)$ we may write 
(\ref{eq12}) in the form
\begin{equation}
\frac{\partial W}{\partial t} = e^{tH}\, (\hat L_{\rm FP} W) e^{-tH},
\label{eq19}
\end{equation}
where 
\begin{equation}
\hat L_{\rm FP} = \frac{\hbar^2}{2} \left[ \nabla^2 
  + \frac{2i}{\hbar} \vec{p}\cdot\nabla \right] .
\label{eq20}
\end{equation}
For (\ref{eq14}) we then obtain
\begin{equation}
\frac{\partial W}{\partial t} = e^{tH}\, (\nabla\cdot\vec{J}) e^{-tH},
\label{eq21}
\end{equation}
with 
\begin{equation}
\vec{J} = \frac{\hbar^2}{2} \left[ \nabla 
  + \frac{2i}{\hbar} \vec{p} \right] W .
\label{eq22}
\end{equation}
Equations (\ref{eq19}) and (\ref{eq21}) are of form of a FPE, just
like eq.~(\ref{eq07}).

Another, equivalent form of the same equation (\ref{eq12}) is easily 
derived:
\begin{equation}
\frac{\partial W}{\partial t} = \frac{1}{2} e^{tH}\, 
  \left( {\hat{\vec{\cal P}}}^2 - \vec{p}^2 \right) e^{-tH},
\label{eq23}
\end{equation}
with the quantum operator
\begin{equation}
{\hat{\vec{\cal P}}} = \vec{p} - i\hbar\nabla .
\label{eq24}
\end{equation}
This is the Bloch-Schr\"odinger form of eq.~(\ref{eq07}).

As a check, from $Z(t) = \int d\Gamma W e^{-tH}$ we have, using 
(\ref{eq23}):
\begin{equation}
\frac{\partial Z}{\partial t} = -\int d\Gamma
  \left(\frac{1}{2}{\hat{\vec{\cal P}}}^2 + V(\vec{x}) \right)
  W\, e^{-tH} .
\label{eq25}
\end{equation}
Integrating over $t$ we have
\begin{equation}
Z(t) = \int d\Gamma \left(\frac{1}{2}{\hat{\vec{\cal P}}}^2 
  + V(\vec{x}) \right) \int_t^\infty dt' e^{-t'H}\, W(\vec{x},\vec{p};t') .
\label{eq26}
\end{equation}
In the limit $\hbar = 0, \hat{\vec{\cal P}} = \vec{p}$, (and thus 
$W=1$) eq.~(\ref{eq26}) reduces to the Thomas-Fermi term
\begin{equation}
Z(t) = \int d\Gamma e^{-tH} .
\label{eq27}
\end{equation}

Finally, we mention another form of the basic equation (\ref{eq07}),
or its equivalent, eq.~(\ref{eq10}). If one introduces instead of $W$
the new function $\tilde W = \exp(-tV)\,W$, then (\ref{eq10}) can be 
written in the form
\begin{equation}
\frac{\partial\tilde W}{\partial t} 
= \left[ \frac{\hbar^2}{2}{\vec{\tilde{\cal D}}}^2 
  + i\hbar \vec{p}\cdot\vec{\tilde{\cal D}} \right] {\tilde W}
  - \frac{1}{2} V {\tilde W} ,
\label{eq28}
\end{equation}
where the operator $\vec{\tilde{\cal D}}$ is given by
\begin{equation}
\vec{\tilde{\cal D}} = e^{-tV/2}\,\nabla\,e^{-tV/2} .
\label{eq29}
\end{equation}
Equation (\ref{eq28}) corresponds to the FPE used in supersymmetric
quantum mechanics in one dimension in the association with the 
Darboux transformation (see, e.~g.~\cite{bib12}).

\section{More on the Bloch equation}

We see that WK semiclassical expansion leads to several equations
with interesting connections to fundamental branches of physics.
The basis of this expansion is the Bloch equation (\ref{eq05}).
Having obtained the solution of (\ref{eq05}) we find the partition 
function 
\begin{equation}
Z(t) = \int d\Gamma e^{-i\vec{p}\cdot\vec{x}/\hbar} u(\vec{x},\vec{p};t) .
\label{eq30}
\end{equation}
The initial condition
\begin{equation}
u(\vec{x},\vec{p};0) = e^{i\vec{p}\cdot\vec{x}/\hbar}
\label{eq31}
\end{equation}
may appear somewhat unusual taking into account that the potentials 
$V(\vec{x})$ considered here are generally associated with the bound 
states of the corresponding Schrodinger equation. Nevertheless,
starting from (\ref{eq05}), we may represent $u(\vec{x},\vec{p};t)$ 
as an expansion in terms of plane waves satisfying the initial 
condition (\ref{eq31}):
\begin{equation}
u(\vec{x},\vec{p};t) = \int d^np' a(\vec{p},\vec{p}';t)
  e^{i\vec{p}'\cdot\vec{x}/\hbar - p'^2t/2} .
\label{eq32}
\end{equation}
Equation (\ref{eq31}) translates into the following initial condition
for the amplitude $a(\vec{p},\vec{p}';t)$:
\begin{equation}
a(\vec{p},\vec{p}';0) = \delta(\vec{p} - \vec{p}') ,
\label{eq33}
\end{equation}
and eq.~(\ref{eq05}) gives, after multiplying the resulting equation
by $\exp(-i\vec{p}\cdot\vec{x}/\hbar)$ and integrating both sides
over $x$:
\begin{eqnarray}
\frac{\partial}{\partial t} \left[a(\vec{p},\vec{p};t) e^{-p^2t/2}\right]
= -\int \frac{d^np'}{(2\pi\hbar)^n} a(\vec{p},\vec{p}';t) e^{-p'^2t/2}
\nonumber \\ \qquad\qquad
  \int d^nx V(\vec{x}) e^{i(\vec{p}' - \vec{p})\cdot\vec{x}/\hbar} .
\label{eq34}
\end{eqnarray}
Introducing the function 
\begin{equation}
A(\vec{p},\vec{p}';t) = a(\vec{p},\vec{p}';t) e^{-p'^2t/2}
\label{eq35}
\end{equation} 
we obtain
\begin{equation}
\frac{\partial}{\partial t} A(\vec{p},\vec{p};t) 
  + \frac{p^2}{2} A(\vec{p},\vec{p};t) 
= -\int d^np' A(\vec{p},\vec{p}';t)  {\tilde V}(\vec{p}-\vec{p}') ,
\label{eq36}
\end{equation}
where
\begin{equation}
{\tilde V}(\vec{p}) = \int \frac{d^nx}{(2\pi\hbar)^n} 
  V(\vec{x}) e^{-i\vec{p}\cdot\vec{x}/\hbar}
\label{eq37}
\end{equation}
Inserting (\ref{eq32}) into eq.~(\ref{eq30}) we find that 
\begin{equation}
Z(t) = \int d^np A(\vec{p},\vec{p};t) .
\label{eq38}
\end{equation}
This form is analogous to the expression for $Z(t)$ in terms of
the Bloch density matrix (see, e.~g.~\cite{bib01}):
\begin{equation}
C(\vec{x},\vec{x}';t) 
= \sum_n \psi_n^*(\vec{x})\psi_n(\vec{x'}) e^{-tE_n}
= \langle \vec{x}| e^{-t\hat H} | \vec{x}' \rangle ,
\label{eq39}
\end{equation}
from which it is obvious that the quantity
\begin{equation}
A(\vec{p},\vec{p}';t) = \langle \vec{p}| e^{-t\hat H} | \vec{p}' \rangle ,
\label{eq43}
\end{equation}
whose trace is the partition function, is the counterpart of 
$C(\vec{x},\vec{x}';t)$ in the momentum representation.

As an example, we consider the case of $n=2$, i.~e.~the Yang-Mills 
quantum mechanics (YMQM) with the potential
\begin{equation}
V(x,y) = \frac{g^2}{2} x^2 y^2 ,
\label{eq44}
\end{equation}
which was the subject of numerous studies (see \cite{bib10} for a review).
This potential leads to a non-separable and non-integrable Schr\"odinger
equation. Its Fourier transform is given by
\begin{equation}
{\tilde V}(\vec{p}'-\vec{p}) = \frac{g^2\hbar^4}{2}
\frac{\partial^2}{\partial {p'}_x^2}\,\frac{\partial^2}{\partial {p'}_y^2}
\delta(\vec{p}'-\vec{p}) ,
\label{eq45}
\end{equation}
which lets (\ref{eq36}) take the form
\begin{equation}
\frac{\partial}{\partial t} A(\vec{p},\vec{p};t) 
  + \frac{p^2}{2} A(\vec{p},\vec{p};t) 
= - \left. \frac{g^2\hbar^4}{2} \frac{\partial^2}{\partial {p'}_x^2}\,
   \frac{\partial^2}{\partial {p'}_y^2} A(\vec{p},\vec{p'};t) 
   \right|_{\vec{p'}=\vec{p}} .
\label{eq46}
\end{equation}
From (\ref{eq46}) one finds that the quantum corrections to the 
classical result for $Z(t)$ generally begin at the order $\hbar^4$
for the YMQM model \cite{bib08}. In the same way, one may show that
the quantum corrections to $Z(t)$ for the harmonic oscillator begin
at the order $\hbar^2$, in agreement with the result found by the
expansion of the well-known exact expression for $Z(t)$. The generalization 
of (\ref{eq46}) to higher dimensions ($n>2$) is straightforward.

\section{Conclusions}

In the present paper we have presented several different forms of the 
differential equation for the quantum corrections to the partition 
function $Z(t)$. Some of these forms appear to be novel. Any one of
these equivalent equations can be used to generate the gradient expansion 
of $Z(t)$ for a wide class of potentials or could serve as the 
starting point for a numerical evaluation. The analogy between some
forms of the equation and other well studied equations of theoretical 
physics, such as the Fokker-Planck equation, suggests the possibility
of novel approaches or approximation schemes for the calculation of 
$Z(t)$. As an example, we have used the integral representation of the
Bloch equation in momentum space to give a simple proof of the fact 
that the quantum corrections to $Z(t)$ for the anharmonic Yang-Mills 
quantum mechanics begins at the order $\hbar^4$ \cite{bib08}. We have
also found a hitherto unknown correspondence with the Fokker-Planck 
equation occurring in supersymmetric quantum mechanics.

\end{document}